\documentclass[a4paper,twocolumn,showpacs,superscriptaddress,floatfix]{revtex4-2}
\usepackage{graphicx}
\usepackage{epsf}
\usepackage{epsfig}
\usepackage{amssymb,amsmath}
\usepackage[usenames]{color}
\usepackage{amssymb}
\usepackage{times}

\newcommand{\be}{\begin{equation}}
\newcommand{\ee}{\end{equation}}
\newcommand{\bea}{\begin{eqnarray}}
\newcommand{\eea}{\end{eqnarray}}
\newcommand{\nn}{\nonumber}

\font\tenscr=rsfs10 scaled1100
\font\sevenscr=rsfs7 
\font\fivescr=rsfs5 
\skewchar\tenscr='177
\skewchar\sevenscr='177
\skewchar\fivescr='177
\newfam\scrfam
\textfont\scrfam=\tenscr
\scriptfont\scrfam=\sevenscr
\scriptscriptfont\scrfam=\fivescr

\begin{document}

\title[Painlev\'e-II approach to binary black hole merger dynamics: universality from integrability]
{Painlev\'e-II approach to binary black hole merger dynamics: universality from integrability }

\author{Jos\'e Luis Jaramillo}

\address{Institut de Math\'ematiques de Bourgogne  (IMB), UMR 5584,  CNRS, \\ Universit\'e  de  Bourgogne
  Franche-Comt\'e,  F-21000  Dijon,  France}

\author{Badri Krishnan}

\address{Institute for Mathematics, Astrophysics and Particle Physics\\
  Radboud University, Heyendaalseweg 135, 6525 AJ Nijmegen, The
  Netherlands}
\address{Max-Planck-Institut f\"ur Gravitationsphysik (Albert Einstein
  Institute)\\ Callinstra{\ss}e 38, 30167 Hannover, Germany}

\vspace{10pt}

\begin{abstract}
The binary black hole merger waveform is both simple and universal. Adopting an
effective asymptotic description of the dynamics, we aim at accounting for such
universality  in terms of underlying (effective) integrable structures. More specifically,
under a ``wave-mean flow'' perspective, we propose that fast degrees of freedom corresponding
to the observed waveform would be subject to effective linear dynamics, propagating
on a slowly evolving background subject to (effective) non-linear integrable dynamics.
The Painlev\'e property of the latter would be implemented in terms of the
so-called Painlev\'e-II transcendent, providing a structural link
between i) orbital (in particular, EMRI) dynamics in the inspiral phase, ii) self-similar solutions of
non-linear dispersive Korteweg-de Vries-like equations (namely, the `modified Korteweg-de Vries' equation)
through the merger and iii) the matching with the isospectral features of black hole quasi-normal modes
in late ringdown dynamics.
Moreover, the Painlev\'e-II equation provides also a `non-linear turning point' problem, extending
the linear discussion in the recently introduced Airy approach to binary black hole merger waveforms.
Under the proposed integrability perspective, the simplicity and universality of the
binary black hole merger waveform would be accounted to by the `hidden symmetries' of the
underlying integrable (effective) dynamics.
In the spirit of asymptotic reasoning, and considering
Ward's conjecture linking integrability and self-dual Yang-Mills structures,
it is tantalizing to question if such universal patterns would reflect the actual
full integrability of a (self-dual) sector of general relativity, ultimately responsible for the  
binary black hole waveform patterns.

\end{abstract}

\pacs{}

\maketitle


\section{Binary black hole mergers: universality and simplicity}

The gravitational binary black hole (BBH) merger waveform is
remarkable simple. This is confirmed by numerical simulations of black
hole mergers within general relativity \cite{Pre05}, and also by
observations of the emitted gravitational radiation
\cite{Abbott:2016blz}.  The inspiral regime can be very complicated
especially if one begins with a complicated initial configuration with
high eccentricity, spins, etc.  However, by the time we reach the
merger, gravitational radiation emission has served to circularize the
orbit and eventually the complexity of the initial configuration is
lost.  This argument suggests that the merger waveform is simple and
in fact also universal.

Such simplicity is consistent with existing treatments of BBH
dynamics, but a clear cut identification of the mechanism(s) behind it
is still missing.  In a companion article \cite{JarKri21}, a
catastrophe-theory model based on the structural stability of fold
caustics --- and of the diffraction patterns upon such caustics--- has
been proposed to effectively account for such universality and
simplicity of the BBH merger waveform. Such a model leads, in
particular, to a phenomenological BBH waveform proposal based on
reparameterized Airy function, in what might be referred to in
\cite{JarKri21} as a `post-Airy expansion' approach.  Thus, the Airy
function serves as a leading order approximation to the merger
waveform, and more accurate approximations can be obtained
perturbatively from this starting point.

The goal here is to advance in the understanding of this problem from
a different perspective.  Whereas the Airy model focused on structural
features related to the phase function of the propagating wave and was
largely (but not completely) independent of the details of the
underlying gravitational dynamics, the present work aims at targeting
the specific structural properties of the BBH gravitational system
responsible of the universality and simplicity of the BBH merger.  In
other words, whereas the Airy model was explicitly meant as an
effective treatment aiming at simplifying the functional modelling of
the waveform with data analysis purposes, the present discussion aims
at unveiling the underlying fundamental mechanisms and structures
behind simplicity and universality in BBH dynamics.

Such effective and fundamental approaches are not in contraposition,
but are indeed complementary. In particular, at a methodological
level, they share an {\em asymptotic reasoning}
strategy~\cite{Batte01} aiming at filtering the overload of details
possibly encumbering structural features in complete approaches. In
such an asymptotic approach a theory is typically described in a
certain limit of a small (or large) parameter, explicitly sacrificing
precision and exactness by eliminating details, in order to make
underlying patterns explicitly apparent in the appropriate range of
the asymptotic parameter, often entailing a gain in the mathematical
tractability and, more importantly in the present context, identifying
at the asymptotic regime the mathematical structure actually present
in the complete dynamics.  In sum, both the effective approach
\cite{JarKri21} and the first-principles here adopt an {\em asymptotic
  reasoning} methodology, but with a focus on different levels of
description.

\subsection{From Airy waveforms to Painlev\'e-II dynamics}
\label{s:from_Airy-PII}
In \cite{JarKri21} a fold-caustic diffraction model on the phase of the
waveform led to an Airy-like model for the BBH waveform.
Specifically, dwelling in a geometric optics setting and adopting the language
of catastrophe theory, 
caustics permit to account for the topological structure underlying the abrupt
illumination-darkness transition in a detector. Then, taking a first step from geometric optics
to wave theory, diffraction on a caustic regularizes the caustic divergence and
provides universal diffraction pattern only depending on the topology of the caustic.
Building on these elements,  we proposed in \cite{JarKri21}
a simple model for the BBH merger waveform based on the assumption of a fold-caustic
structure in-built in the phase-delay function describing the gravitation wave propagation,
identifying the Airy function as the reference functional form to capture
the qualitative features of the waveform merger.
Being based on
a first order correction to geometric optics,
such a treatment is explicitly a high-frequency one. It was only in the
{\em asymptotic reasoning} spirit that such an Airy-like pattern was
claimed in \cite{JarKri21} to extend to all frequencies.

In this article we provide a more robust justification of such extension of the Airy-like
model to arbitrary BBH waveform frequencies.
Building on the solid catastrophe-theory identification of the Airy function
in the high-frequency limit, we address the extension of the phenomenological model
to all frequencies by adopting the following simple assumption:  the BBH waveform smoothly
(and linearly) transitions from an oscillatory regime to a
damped regime. Adopting the language of (semi-)classical treatments of dynamics,
such  dynamical transitions from oscillatory to damped regimes
define a so-called `turning point' problem.
The Airy equation
\bea
\label{e:Airy_equation}
\frac{d^2u}{d\tau^2}  - \tau u = 0 \ ,
\eea
provides precisely the universal leading-order of linear turning points. 
Here $\tau$ is a reparameterized time parameter.  
In section \ref{s:linear_turning_points} we recover the full
functional dependence of BBH merger waveform derived in
caustic diffraction model, through a `turning point' ordinary differential equation (ODE)
reasoning not restricted to high frequencies.

The methodology described above can indeed be seen as another instance
of asymptotic reasoning in which, instead of simply extending the
high-frequency result to all frequencies, we have rather identified
the fundamental analytical {\em structure} underlying the Airy
function, namely the Airy equation, and promoted it as the guiding
element to extend the phenomenological Airy description of the BBH
waveform to {\em all} frequencies.  This more abstract version of
asymptotic reasoning, in which it is the underlying fundamental
structure and not the waveform functional form itself, leads to the
prototypical argument that we shall follow.

Actually, the ODE `turning point' change of perspective has profound implications.
Indeed, the path to the Airy equation \eqref{e:Airy_equation} relies strongly
on the assumption of linearity of waveform, supported by accumulated insights
\cite{Price:1994pm,Baker:2001sf,Buonanno:2006ui,Berti:2007fi,Giesler:2019uxc}).
But we can refine this notion. Let us consider that:
\begin{itemize}
\item[i)] The effectively linearity of the
 propagating wave occurs on an effective background whose dynamics are
 indeed non-linear. This provides an effective separation between fast (linear propagating wave) and
 slow (non-linear background) degrees of freedom.

\item[ii)] The actual `turning point' dynamics happens at the level
of nonlinear background dynamics and, only at a second stage, are imprinted in the propagating
linear wave.
\end{itemize}
In this perspective, the Airy equation (\ref{e:Airy_equation}) would represent
the linear imprint in the propagating wave  of the underlying background `turning point'
dynamics. The question of a non-linear version of `turning point' dynamics is naturally posed.

If the Airy equation (\ref{e:Airy_equation}) represents the ``archetype'' of linear turning problem,
the so-called Painlev\'e-II equation, namely
\bea
\label{e:Painleve-II_complete}
\frac{d^2 w}{d\tau^2} -\tau w -2 w^3 - \alpha = 0 \ ,
\eea
with $\alpha$ an arbitrary constant, can be seen  as an archetype of ``non-linear turning point''~\cite{AblSeg77}.
In the same way that Eq. (\ref{e:Airy_equation}) defines (under the appropriate
asymptotic conditions) the Airy special function, Eq. (\ref{e:Painleve-II_complete}) defines
a new transcendental special function $w(\tau)$ known as the  Painlev\'e-II transcendent
(note that, in full generality, $\tau$ must be seen as a complex variable).
This equation reduces upon linearization  (in the case  $\alpha=0$) to the Airy one, and 
the Painlev\'e-II transcendent has asymptotic behaviour controlled indeed by the Airy function
(see section \ref{s:Painleve_transcendents} for a better assessment of this statement).
This special function, together with the other five Painlev\'e transcendents
(see e.g. \cite{Clark03}), plays a prominent role in mathematical physics,
in particular in connection with integrable systems. This immediately prompts
a new perspective on univesality and simplicity of BBH mergers.
Specifically, once adopted an non-linear ODE `turning point' perspective to the BBH merger
dynamics, the following question is posed: {\em could  non-linear Painlev\'e transcendents,
and more specifically Painlev\'e-II, account for the observed particular features of the BBH dynamics?}

The previous question is, admittedly, a bold one, and can be
considered only as a conjecture for now. The remarkable
fact is that it is indeed supported by existing results.  A first
instance is presented in a notable article by
Rajeev~\cite{Rajeev:2008sw}, in the setting of the damped orbital
motion of a charged particle in a Coulomb potential, subject to
radiation reaction.  Specifically, Rajeev shows that the
Landau-Lifshitz equations for a charged particle moving in a Coulomb
potential can be solved exactly in terms of the Painlev\'e-II equation
with $\alpha=0$ \bea
\label{e:Painleve-II_alpha=0}
\frac{d^2 w}{d\tau^2} -\tau w -2 w^3  = 0 \ .
\eea
Moving now to general relativity itself, consider the extreme mass ratio case
where we have a small compact object orbiting around a large black hole. The motion of the small object can be
viewed as a sequence of slowly evolving geodesics, suffering the effects of radiation damping, and losing energy
and angular momentum in the form of gravitational radiation. However, when it gets sufficiently close to the
large black hole, the motion transitions to a plunge. This picture was suggested by Thorne and Ori in Ref.
\cite{Ori:2000zn}. In the inspiral regime, the motion is governed by an effective potential $V(r, L)$ with
$r$ being a radial coordinate and $L$ the angular momentum. This potential has a minimum in the inspiral regime
but as $L$ decreases, the minimum changes to a saddle point and then disappears. Thorne and Ori showed that in
this transition regime, the dynamics is governed by an  equation that, as noticed by Comp\`ere and K\"uchler
\cite{Compere:2021iwh,Compere:2021zfj}, can be rewritten as
\bea
\label{e:Painleve-I}
\frac{d^2 w}{d\tau^2} - \tau -6w^2 = 0 \ ,
\eea
namely the Painlev\'e-I equation. The relevant Painlev\'e-I transcendent in
the plunge problem is, in particular, the so-called {\em tritronqu\'ee} solution to Painlevé-I 
\cite{DubGraKle09}, a function associated with universality in critical phenomena
of certain partial differential equations, a remarkable feature by itself.
On the other hand, it can be shown that the Painlev\'e-I equation (\ref{e:Painleve-I}) can be obtained as a
(singular) limit of Painlev\'e-II (\ref{e:Painleve-II_complete}).
In view of these elements, it becomes plausible that 
the orbital dynamics is ultimately governed by the second Painlevé equation. This will discussed
in section \ref{e:EMRI-PII}.

A natural criticism to the previous discussion refers to its restriction to
the extreme mass ratio inspiral (EMRI) limit of the BBH dynamics.
This is indeed the case and we claim that it is precisely the focus
on this asymptotic limit of vanishing mass ratio, that allows to
identify ---as a new instance of asymptotic reasoning-- the relevant
underlying pattern in the full dynamics, namely: {\em the fundamental structural
  role of Painlev\'e transcendents ---and in particular of Painlev\'e-II--- in BBH dynamics,
  supports the study of BBH mergers in the setting of integrable or quasi-integrable systems}.

\subsection{Universal wave patterns: from ODEs to PDEs}
Painlev\'e transcendents are naturally introduced in an ODE setting,
as in the `turning point' problems discussed above. However, their
structural role transcends ODEs, extending to very different problems
threaded by the notion of integrability.
In our particular BBH problem, if the inspiral phase (at least,
up to the plunge) admits natural treatments in terms of ODEs, the situation
for the merger and ringdown phases is less clear and a partial differential equation
(PDE) treatment could be better adapted.
Remarkably, Painlev\'e transcendents provide a structural link
to an important class of non-linear integrable PDEs with a key
role in the integrability properties of BHs.

\subsubsection{Universal wave patterns in dispersive non-linear PDEs}
  Before tackling the discussion of Painlev\'e transcendents in
  BBH mergers, let us make an interlude into dispersive non-linear hydrodynamics,
  specifically in the context of dispersive shock waves \cite{Mille16}.

  A remarkable universality feature in this setting is that {\em all shocks look the same}~
  \footnote{We acknowledge P. Miller for calling our attention to the example of Burgers'
    shock.}. 
Specifically, let us we consider the Burgers' equation with  viscosity term regularization
\bea
\label{e:Burgers_epsilon}
   \partial_t u + u\partial_x u = \epsilon \partial^2_{xx} u \ .
   \eea
   Inviscid Burgers equation ($\epsilon=0$) with initial data $u_0(x)=u(0,t)$
   develops a shock propagating along a curve $x=x_c(t)$ depending on $u_0(x)$.
   If we denote $u_-(t)= \lim_{x\to-\infty} u(x, t)$, $u_+(t)= \lim_{x\to-\infty} u(x, t)$,
   and we zoom around the ``critical'' shock position $x_c(t)$ in the solution to the regularized
   equation (\ref{e:Burgers_epsilon}), one finds 
   the solution is (e.g. \cite{LafOMa95,GooXin92})
   \bea
\label{e:universal_shock}
u(\xi,t) &=& \lim_{\epsilon\to 0}  u(x_c(t)+\epsilon \xi,t) \\
&=& \bar{u}(t) - \frac{\Delta u(t)}{2}\tanh\left(\frac{(\xi - \xi_0(t))\Delta u(t)}{4} \right)
\ , \nn
\eea
where $\bar{u}(t) = (u_-(t)+u_+(t)/2$, $\Delta u(t)= u_-(t)-u_+(t)$
and $\xi_0(t)$ is a function that can be fixed at higher orders.
The remarkable facts are that, in ``appropriate'' coordinates:
\begin{itemize}
\item[i)] The form is fully independent of the
initial data $u_0(x)$.
\item[ii)] Such form is given by a universal special function,
  specifically  the (elementary) special function $\tanh(x)$.
\end{itemize}
Initial data enter through $x_c(t), u_-(t), u_+(t)$ and $ \xi_0(t)$
in the reparametrization of the function describing the shock.

One could think that such wave pattern universality
is a very peculiar property of shocks, but actually it is
not. Indeed, 
such universality is a generic feature associated
with a PDE-type of critical behaviour~\cite{Mille16}
occurring at transitional dynamical wave patterns in integrable dispersive equations~\footnote{Actually, 
according to Dubrovin's conjecture~\cite{Dubrovin2006},
such universility of PDE critical behaviour extends beyond
integrabililty to generic Hamiltonian perturbations of hyperbolic
systems.}.

\subsubsection{Painlev\'e-II and universal BBH mergers}
At this point we come back to the Painlev\'e-II transcendent,
identified as a key structure of BBH dynamics in the ODE inspiral phase,
but now from a (merger) PDE perspective.

In the light of discussion above on universal wave patterns,
we explore the possibility that the observed universality
of BBH merger dynamics could respond to a PDE-type
of critical behaviour. More specifically, we consider if 
non-linear background BBH dynamics can be 
encoded in an effective PDE ---analogous role to  Burgers' Eq. (\ref{e:Burgers_epsilon})---
such that the BBH merger transient ---analogue to the  Burgers' shock transient---
can be understood in terms of a
PDE-type critical behaviour with universal dynamics encoded
in the Painlev\'e-II transcendent ---special function analogous to $\tanh(x)$ in Eq. (\ref{e:universal_shock}).

In section \ref{s:BBH_PDE_integrability} we will explore this possibility under
two radical assumptions: i) {\em dispersive nature} of the effective background non-linear
dynamics, and ii) {\em self-similar character} of the solution at the (``critical'')
merger transition. The first assumption is directed motivated by the
Airy-BBH model in \cite{JarKri21} and the fact that Airy function controls as
the generic first-order behaviour of all dispersive systems.
Ultimately, such dispersive character should
be justified in terms of an (asymptotic) coarse-grained PDE description
where integration over small scales would result on an effective dispersion.
Regarding the self-similar assumption, it is motivated by the
assumed  critical behaviour at the merger transient \cite{barenblatt_1996}.

Regarding the non-linear dispersive Ansatz, a natural ``universal'' PDE model 
is the Korteweg-de Vries (KdV) equation
\bea
\label{e:KdV}
\partial_t u + 6 u\partial_xu + \partial^3_{xxx}u = 0 \ .
\eea
More specifically,  KdV-type equations give the leading-order asymptotic equations
of weakly dispersive and weakly non-linear PDE dynamics. 
In this setting, and in order to incorporate the second (self-similarity) assumption
in connection with the Painlev\'e-II transcendent, the natural
candidate is the so-called modified-Korteweg-de Vries (m-KdV)
equation~\footnote{As an interesting synergy,
  Damour and Smilga \cite{Damour:2021fva}
  have identified precisely the  Painlev\'e-II transcendent as
  the one instances in a classification of  ghost-free Hamiltonian systems,
  providing also the connection with modified Korteweg-de Vries. This
  synergy is remarkable by itself.}
\bea
\label{e:modified_KdV}
u_t - 6u^2\partial_xu + \partial^3_{xxx}u = 0 \ ,
\eea
related to the standard KdV Eq. (\ref{e:KdV})  by an appropriate B\"acklund transformation.
Our claim is that there exists a coordinate system in 
which the effective dispersive non-linear background BBH dynamics
is described by the mKdV equation.
Under the appropriate self-similarity assumption, Eq. (\ref{e:modified_KdV})
can be solved in terms of the Painlev\'e-II transcendent \cite{AblSeg77}. This is
an instance of the so-called ``Painlev\'e tests'', characterizing integrable
PDEs through their solvability in terms of Painlev\'e transcendents in
an inverse scattering transform (IST) scheme~\cite{Gelfa51,March55}.
This scattering perspective connects with the mathematical
framework of the late ringdown phase,
naturally described in a (direct) scattering theory setting~\cite{Chandrasekhar:579245}.

In sum, we propose Painleve-II transcendent as a
fundamental underlying structure in BBH dynamics, cast
in two complementary realizations: a) the ODE controlling the inspiral and transition
to the plunge (in its Painlev\'e-I limit), and b) the universal
PDE wave pattern in the merger transient to the ringdown.
Painleve-II is therefore the key guideline to
approach BBH mergers as (quasi-)integrable systems.

\bigskip

The plan of the article is the following. In section \ref{s:turning_points}
we revisit the Airy-function approach to BBH merger waveform,
taking an `turning point' ODE approach that permits to extend
results in~\cite{JarKri21} to all frequencies. Once in the
ODE setting, in section~\ref{s:Painleve_ODE} we start by reviewing
of Painlev\'e transcendents and Painlev\'eRajeev's
solution of radiation-reaction charged particle orbital motion in terms
of Painlev\'e-II as well as, in a gravitational setting, the emergence
of Painlev\'e-I in the transition to the plunge of EMRIs. Then  we
extend Rajeev's discussion to the gravitational radiation reaction setting,
in the setting of quasi-circular orbits, leading to the identification
of Painlev\'e-II as a fundamental structure in BBH merger dynamics.
Section~\ref{s:BBH_PDE_integrability} has a more heuristic flavor,
in which a separation of fast and slow degrees is proposed, in such
a way that slow background BBH dynamics is approached in terms of a
non-linear dispersive integrable equation solvable in terms of
the Painlev\'e-II transcendent through n inverse
scattering transform. This configures a self-contained
bottom-up approach to the BBH merger dynamics in terms
of a (quasi-)integrable system determined by  Painlev\'e-II.
We conclude in section~\ref{s:integrability_background_dynamics} by considering
integrability from a larger scope as a guiding principle to
match our `bottom-up' approach to BBH dynamics and waveform with a `top-down'
scheme to be developed.

\section{BBH 'turning point' models: from Airy to Painlev\'e-II}
\label{s:turning_points}

\subsection{Linear oscillation-to-damped transition: Airy case}
\label{s:linear_turning_points}
      In ref.~\cite{JarKri21} we adopted an {\em asymptotic reasoning} approach to extend the
      the Airy model for the BBH waveform merger, based on the diffraction on caustics and valid
      for high frequencies, to all frequencies.
      Let us take here rather an ``agnostic'' perspective regarding the underlying mechanism driving the BBH merger
dynamics and focus on the general structural fact that this dynamics describe a 
dynamical transition from an oscillatory regime of the system to a regime with damped sinusoids
in a continuous and smooth way.

In semiclassical treatments of dynamics such transitions are related to so-called
``turning points''. The Airy equation
\bea
\label{e:Airy_equation}
\frac{d^2u}{d\tau^2}  - \tau u = 0 \ ,
\eea
is the archetype of a linear turning point. 
We focus here on a more general class of  linear ODEs describing, in an effective way, a turning point  problem. 
Namely, we consider equations of the form~\cite{wasow2018asymptotic}
\bea
\label{e:turning_point_equation}
\epsilon^2 \frac{d^2u}{d\tau^2} - \left(\tau \phi(\tau)  - \epsilon\psi(\tau,\epsilon) \right)u = 0 \ \ , \ \ \phi(0)\neq 0 \ ,
\eea
where $\epsilon$ is a small parameter and $\phi(\tau)$ and $\psi(\tau,\epsilon)$ are given analytic
functions,
encoding in an effective manner the physical features of the turning point problem. 
The Airy equation is recovered by reparametrizing $\tau\to \tau/\epsilon$ and setting
$\phi(\tau)=\epsilon$ and $\psi(\tau,\epsilon)=0$. 
We perform a first order reduction of (\ref{e:turning_point_equation}) by introducing the variable
\bea
Y =
\left(
\begin{array}{c}
   u \\
 \epsilon \dot{u}
\end{array}
\right)
\eea
where $\dot{u} = \frac{du}{d\tau}$, so that Eq. (\ref{e:turning_point_equation}) writes
\bea
\label{e:turning_point_1st_order}
\epsilon \dot{Y}
=
\left(
  \begin{array}{c|c}
    0 & 1 \\
    \hline 
    \tau \phi(\tau) + \epsilon\psi(\tau,\epsilon) & 0  
  \end{array}
  \right) Y \equiv A(\tau, \epsilon) Y \ ,
\eea
this expression defining the matrix $A(\tau, \epsilon)$. We note that in the
Airy case
\bea
\label{e:matrix_A_Airy}
A_\mathrm{Airy}(\tau) 
\left(
  \begin{array}{c|c}
    0 & 1 \\
    \hline 
    \tau & 0  
  \end{array}
  \right) \ .
\eea
Solutions to Eq. (\ref{e:turning_point_1st_order}) can indeed 
be expressed in terms of Airy functions solutions to Eq. (\ref{e:Airy_equation}).
Specifically, Theorem 29.1 in  ~\cite{wasow2018asymptotic} states that, given 
a $2\times2$ matrix $A(\tau, \epsilon)$ as defined in (\ref{e:turning_point_1st_order}),
holomorphic 
in a neighborhood of $(0,0)$ and such that
\begin{itemize}
\item[i)] $A(0,0)$ is similar to the matrix  
\bea
\left(
  \begin{array}{c|c}
    0 & 1 \\
    \hline 
   0  & 0  
  \end{array}
  \right) \ ,
\eea
\item[ii)] It holds
\bea
\frac{d}{d\tau} \mathrm{det} A(\tau, 0)|_{\tau=0} \neq 0 \ ,
\eea
\end{itemize}
 then solutions to the "turning point" problem of the form
\bea
\label{e:turning_point_A}
\epsilon \dot{Y}(\tau) = A(\tau, \epsilon) Y(\tau)  \ ,
\eea
can be written as
\bea
\label{e:sol_Ai_Aiprime}
Y(\tau) = P(\tau) e^{\frac{1}{2\epsilon}\int_0^\tau \mathrm{Tr}(A(\lambda,0)d\lambda} \tilde{Y}(t) \ ,
\eea
where $t=\Phi(\tau)$, with\footnote{$\Phi$ is denoted $\alpha$ in \cite{wasow2018asymptotic}. } $\Phi(\tau)$
holomorphic at $\tau=0$ with $\Phi(0)=0$ and
$\dot{\Phi}(0)\neq 0$, and $P(\tau)$ is a holomorphic matrix function (with inverse also holomorphic) at
$\tau=0$, and $\tilde{Y}(t)$ satisfies the "instantaneous" Airy equation
\bea
\label{e:turning_point_A}
\epsilon \frac{d}{dt}\tilde{Y}(\tau) = \tilde{A}(t, \epsilon) \tilde{Y}(t)  \ ,
\eea
with
\bea
\tilde{A} (t, 0) = A_\mathrm{Airy}(t) =
\left(
  \begin{array}{c|c}
    0 & 1 \\
    \hline 
   t  & 0  
  \end{array}
  \right)
\eea
This shows that solutions to these linear turning point problems are generically
linear combinations of the Airy function $\mathrm{Ai}(t)$ and its derivative
$\mathrm{Ai}'(t)$, with appropriate modulations 
and argument reparametrizations depending on the specific details of the problem, effectively
captured by functions $\phi$ and $\psi$. This matches precisely the structure
found in ref.~\cite{JarKri21}, but without the restriction to high frequencies.

The dynamical transition from oscillating to damped behavior is
captured by the asymptotic behaviour of the Airy function (for
real argument), that we recall here for later comparison 
\bea
\label{e:Airy_asymptotics}
\begin{array}{rcll}
  \displaystyle
 \mathrm{Ai}(t) &\sim&  \frac{1}{\sqrt{\pi}|t|^{\frac{1}{4}}}\sin\left(\frac{2}{3}|t|^{\frac{3}{2}} +
 \frac{\pi}{4}\right) \ \ & (t\to -\infty) \\
 \mathrm{Ai}(t) &\sim& \frac{1}{2\sqrt{\pi}t^{\frac{1}{4}}} e^{-\frac{2}{3}t^{\frac{3}{2}}}
 & (t\to +\infty) \ .
 \end{array}
 \eea
 Note that the overexponential damping at $t\to +\infty$ precludes this linear turning
 point problem from accounting for the late ringdown exponential decay. Another
 mechanism is needed for the transition to the ringdown.

\subsection{Non-linear Airy: Painlev\'e-II 'turning point' model}
\label{s:P_II_turning_point_model}
As commented in \ref{s:from_Airy-PII}, the Painlev\'e-II equation
(\ref{e:Painleve-II_complete}) and, more specifically, the
$\alpha=0$ case (\ref{e:Painleve-II_alpha=0}), provides the
archetype of non-linear turning point equation that reduces
to the Airy discussed above case upon linearization.
A key question is whether the non-linear equation (\ref{e:Painleve-II_alpha=0}) has 
indeed solutions bounded for all real $t$. Then, it is crucial
to understand how to connect the behaviour of this solution at $t\to-\infty$ to that
at $t\to\infty$, to proceed to the ``connection problem''
in a non-linear analogue to the WKB ``connection formulae'' derived from
(\ref{e:Airy_asymptotics}) in the linear Airy case. This requires
a control of the asymptotics of this particular Painlev\'e-II transcendent.

Following  \cite{AblSeg77,HasMcLeo80,SegAbl81,Clark03}, the
asymptotics of the solution to (\ref{e:Painleve-II_alpha=0}) with
boundary condition
\bea
\lim_{t\to\infty} w(t) = 0
\eea
is
\bea
\label{e:P-II_asymptotics_1}
\begin{array}{rcll}
  \displaystyle
 w(t)  &\sim&   \frac{d}{|t|^{\frac{1}{4}}}\sin\left(\frac{2}{3}|t|^{\frac{3}{2}} -
 \frac{3}{4}d^2\ln(|t|) - \theta_o \right)
 & (t\to -\infty) \\
  w(t) &\sim&  \gamma \mathrm{Ai}(t) \sim
  \frac{\gamma}{2\sqrt{\pi}t^{\frac{1}{4}}} e^{-\frac{2}{3}t^{\frac{3}{2}}}
 & (t\to +\infty) 
 \end{array}
\eea
where, given the constant $\gamma$ with $|\gamma|<1$ (for $\gamma\geq 1$ the solution to
(\ref{e:Painleve-II_alpha=0}) diverges at some intermediate point), the value
of amplitude $d$ and the phase shift $\theta_o$ are given (``connection formulae'') by
\bea
\label{e:P-II_asymptotics_2}
d^2(\gamma) &=& -\frac{1}{\pi}\ln(1-\gamma^2)   \\
\theta_0(\gamma) &=& \frac{3}{2}d^2\ln 2 + \mathrm{arg}\left(\Gamma\left(1- \frac{i}{2}d^2\right)\right) - \frac{\pi}{4}    \nn \ .
\eea
It is interesting to compare the Airy asymptotics (\ref{e:Airy_asymptotics})
with Painlev\' e-II ones in  (\ref{e:P-II_asymptotics_1}) and (\ref{e:P-II_asymptotics_2}).
The non-linear turning point introduces both an asymptotique amplitude and phase shift
with respect to the Airy case. In particular, the phase shift 
a corrects the standard $e^{i\pi/4}$ WKB shift, something of potential relevance in the
analysis of the BBH merger waveform.

\section{Painlev\'e transcendents in damped binary EMRI dynamics}
\label{s:Painleve_ODE}

\subsection{Painlev\'e transcendents}
\label{s:Painleve_transcendents}
Apart from providing, as discussed in the previous section, a non-linear generalization of the
turning-point problem defined by the linear Airy equation (\ref{e:Airy_equation}),
solutions to the Painlev\'e-II equation (\ref{e:Painleve-II_complete} displays
remarkable and fascinating mathematical properties shared with a finite number of
other `Painlev\'e transcendents', namely solutions to the following
six Painlev\'e equations (respectively $\mathrm{P_I}-\mathrm{P_{VI}}$)
\bea
\label{e:Painleve_equations}
\frac{d^2w}{dz^2} &=& 6w^2 + z   \\
\frac{d^2w}{dz^2} &=& 2w^3 + zw + \alpha \nn \\
\frac{d^2w}{dz^2} &=& \frac{1}{w}\left(\frac{dw}{dz}\right)^2
-  \frac{1}{z}\frac{dw}{dz} +\frac{\alpha w^2 + \beta}{z} + \gamma w^3 + \frac{\delta}{w} \nn \\
\frac{d^2w}{dz^2} &=& \frac{1}{2w}\left(\frac{dw}{dz}\right)^2
+ \frac{3}{2}w^3 + 4zw^2 + 2(z^2-\alpha)w + \frac{\beta}{w} \nn \\
\frac{d^2w}{dz^2} &=& \left(\frac{1}{2w} + \frac{1}{w-1}\right)\left(\frac{dw}{dz}\right)^2
-\frac{1}{z}\frac{dw}{dz} \nn \\
&+& \frac{(w-1)^2}{z^2}\left(\alpha w +\frac{\beta}{w}\right) +
\frac{\gamma w}{z} +\frac{\delta w(w+1)}{w-1} \nn \\
\frac{d^2w}{dz^2} &=& \frac{1}{2}\left(\frac{1}{w}+\frac{1}{w-1}+  \frac{1}{w-z}\right)
\left(\frac{dw}{dz}\right)^2 \nn \\
&-& \left(\frac{1}{z}+\frac{1}{z-1}+  \frac{1}{w-z}\right)
\frac{dw}{dz} \nn \\
&+& \frac{w(w-1)(w-z)}{z^2(z-1)^2}\left(\alpha + \frac{\beta z}{w^2}
+ \frac{\gamma(z-1)}{(w-1)^2} + \frac{\delta z(z-1)}{(w-z)^2} \right) \nn
\eea
where $\alpha, \beta, \gamma$ and $\delta$ are constants.  A non-linear ODE
is said to possess the so-called `Painlev\'e property'~\footnote{Necessary (in general
  not sufficient) conditions for
an ODE to satisfy the Painlev\'e property
are given by the so called algoritm named as `Painlev\'e test'.
We refer the reader to expositions in \cite{ConMus08,Conte99}.} if the only
`movable' singularity are poles. A singularity of a solution of a given
ODE is `movable' if it is not a priori fixed by the equation
itself, so its location is fixed by initial data. The remarkable
fact is that the Painlev\'e property~\footnote{In our BBH setting,
one can speculate about the possibility of using the `non-movable' property of
the singularities to build on the  universality of the solutions,
since it indicates an indepedence on the initial conditions.
This is a point to be further explored.} severely constraints
the form of the equation, so it can be reduced to one of fifty types of
equations \cite{Ince44,Clark03,AblFok03}, all of them integrable
in terms of known functions except six: precisely the Painlev\'e
equations $\mathrm{P_I}-\mathrm{P_{VI}}$ in  (\ref{e:Painleve_equations}),
which are irreducible to classical special functions and whose
solutions define the Painlev\'e transcendents. The latter
can be seen as a new class of non-linear special functions~\cite{Clark03}.

The relevance of equations satisfying the Painlev\'e properties is that
they can (often) be linearized or solved exactly. Specifically,
for Painlev\'e equations can be studied by using the so-called
isomonodromy method (see references in \cite{Clark03} and, in this
sense, they are understood as integrable equations. This simplicity
and relation to integrability will be the key point in our later
developments. For a recent review of the rich mathematical structures
underlying the Painlev\'e equations, see reference  \cite{Clark19}.

\subsection{Charged particle in Coulomb potential with radiation damping: Painlev\'e-II }
\label{e:Rajeev}
The first contact between damped orbital dynamics and the Painlev\'e transcendents
is presented in Rajeev's remarkable article \cite{Rajeev:2008sw},
in the setting of the orbital motion of a charged under radiation
damping.  Specifically, Rajeev shows that the 
radiation reaction Landau-Lifshitz equations
for a charged particle moving in a Coulomb potential 
including radiation damping  can be solved exactly,  in the
non-relativitic limit, 
in terms of the Painlev\' e-II  equation.

The Landau-Lifschitz equation sof motion of a radiating charged particle in an electrostatic
field can be written in the non-relativistic limit as
\bea
\label{e:LL_eqs_Rajeev}
\frac{d}{dt} \left(\vec{v} + \tau \nabla U\right) + \nabla U = 0 \ ,
\eea
with $U=-\frac{k}{r}$ with $k=q/m$.

Let us revisit Rajeev's argument  in  \cite{Rajeev:2008sw}by slightly extending its discussion in order to better
seize the similarities and differences between the electromagnetic and gravitational cases.
Let us start considering the equation 
\bea
\label{e:LL_eqs}
\frac{d}{dt} \left(\vec{v} + \tau \nabla W\right) + \nabla U = 0 \ ,
\eea
where $\displaystyle \vec{v}=\frac{d\vec{r}}{dt}$, $U=U(r)$ and $W=W(r)$ are radial potentials and $\tau$ is
a constant damping time parameter.
This equation should be understood as the leading (linear) order
in an asymptotic expansion in $\tau$ of the radiation-reaction dynamics.
That is, higher-order powers in $\tau$ are neglected in the analysis.
In particular, the non-relativistic  Landau-Lifshitz equations for a charged
particle in an electrostatic field in \cite{Rajeev:2008sw} are recovered for $\displaystyle U(r)=W(r)=\frac{k}{r}$.

In a first stage, rewriting (\ref{e:LL_eqs}) as
\bea
\frac{d}{dt} \left(\vec{v} + \tau \hat{r} W_r\right) + \hat{r} U = 0 \ ,
\eea
and taking the vector (wedge) product of the equation with $\vec{r}$, we can write
\bea
\label{e:evolution_L}
\frac{d\vec{L}}{dt} = -\tau\frac{W_r}{r} \vec{L} \ ,
\eea
where $\vec{L} = \vec{r}\wedge \vec{v}$ is the angular momentum. In particular, we note that in this approximation the orbital
plane does not change.

In a second stage we use (\ref{e:evolution_L}) to rewrite (\ref{e:LL_eqs}) as (use
standard identities
(8-10) in \cite{Rajeev:2008sw}, and the notation $L=|\vec{L}|$)
\bea
\label{e:ODE_v1}
\frac{dr}{dt} &=& v_r \nn \\
\frac{d}{dt}\left(v_r + \tau W_r\right) &=&  \frac{L^2}{r^3} - U_r \nn \\
\frac{dL}{dt} &=&  -\tau\frac{W_r}{r} L \ .
\eea
Motivated by Rajeev's electrostatic case and the gravitational setting discussed below, let us consider
\bea
\label{e:potentials}
U(r) = \frac{\alpha}{r^n} \ \ , \ \ W(r) = \frac{\beta}{r^m} \ .
\eea
Then, the last equation in (\ref{e:ODE_v1}) becomes
\bea
\frac{dL}{dt} = \frac{m\beta\tau}{r^{m-1}} \frac{L}{r^3} \ ,
\eea
and we can write 
\bea
\frac{L^2}{r^3} = \frac{1}{2\beta m \tau}\left(\frac{d}{dt}\left(r^{m-1}L^2\right) - (m-1)L^2r^{m-2}\dot{r}\right) \ ,
\eea
so that the second equation in (\ref{e:ODE_v1}) writes
\bea
\frac{d}{dt}\left(v_r -\frac{m\beta\tau}{r^{m+1}} -\frac{1}{2\beta m \tau} r^{m-1}L^2 \right) = \frac{n\alpha}{r^{n+1}} - \Delta \ ,
\eea
where 
\bea
\label{e:Delta}
 \Delta = \frac{m-1}{2\beta m \tau}r^{m-2}L^2\dot{r} \ .
\eea
Defining $z$ as
\bea
z = v_r -\frac{m\beta\tau}{r^{m+1}} -\frac{1}{2\beta m \tau} r^{m-1}L^2  \  ,
\eea
we can therefore rewrite the dynamical equations as
\bea
\label{e:ODE_v2}
\frac{dL}{dt} &=& \frac{\beta m \tau}{r^{m+2}}L \nn \\
\frac{dz}{dt} &=& \frac{n\alpha}{r^{n+1}} - \Delta \nn \\
\frac{dr}{dt} &=& \frac{r^{m-1}}{2\beta m \tau}L^2 + z + \frac{m\beta\tau}{r^{m+1}} \ .
\eea
At this point,  the remarkable result of Rajeev is recovered by applying this to the electrostatic case, 
that is, by choosing $\alpha=\beta=k$ and $m=n=1$.
On the one hand, $\Delta = 0$ and eliminating the derivative in time, the electrostatic case can be written as
\bea
\label{e:ODE_Elec}
\frac{dL}{dz} &=& \frac{\tau}{r}L \nn \\
\frac{dr}{dz} &=& \frac{r^2}{2k^2\tau}L^2 + \frac{r^2}{k}z + \tau \ .
\eea
Finally, this leads to
\bea
\label{e:PII_Elec}
\frac{d^2L}{dz^2} = -\frac{1}{2k^2}L^3 -\frac{\tau}{k}z L \ ,
\eea
namely the Painlev\'e-II equation, up to rescaling of $L$ and $z$. We note that
no approximations have been made, once (\ref{e:LL_eqs}) has been adopted.

\subsection{EMRI transition to plunge: Painlev\'e-I dynamics}
\label{e:EMRI-PI}
After discussing the electrostatic case, we proceed to the gravitational setting by 
considering the limit of extreme mass ratio limit of binaries (EMRIs). 
Before adapting Rajeev's argument to a gravitational setting, let us 
comment on other remarkable result on the connection 
between Painlev\' e transcendents and radiation damping 
orbital problems, namely the identification  by Comp\`ere and K\"uchler
\cite{Compere:2021iwh,Compere:2021zfj} of the role of Painlev\'e-I in the plunge equations
originally derived by Ori and Thorne in~\cite{Ori:2000zn}.

Specifically, in \cite{Ori:2000zn} describes the dynamics of this
system in a `transition regime' near the innermost stable circular orbit (ISCO),
between the `adiabatic inspiral regime'
and the `plunge regime'. The radial dynamics is then 
modelled by the geodesic equation corrected with a radial self-force, in a first stage.
However, in a second stage the radial self-force is dropped under the
heuristic justification of only involving a shift in the parameters
of the effective potential.
Adopting some appropriate dimensionless
coordinates $(X, T)$, respectively encoding the radial distance to the
ISCO and the proper time, the dynamics is (cf. Eq. (3.22) in \cite{Ori:2000zn})
\bea
\label{e:Ori_Thorne}
\frac{d^2X}{dT^2} = -X^2 - T \ .
\eea
with asymptotic 
In a remarkable work, Comp\`ere and K\"uchler \cite{Compere:2021iwh,Compere:2021zfj}
revisit the problem with two fundamental contributions: i) justifying
that Eq. (\ref{e:Ori_Thorne}) actually holds as such even when
all effects are systematically taken into account, and ii) identifying
(\ref{e:Ori_Thorne}) as the Painlev\'e-I equation.

Indeed, the latter point connecting to Painlev\'e-I   is justified through a rescaling of $X$ and $T$,
namely introducing
\bea
\label{e:PI_rescaling}
X = \alpha w \ \ , \ \ T = \beta t \ ,
\eea
the choice
\bea
\label{e:PI_rescaling_values}
\alpha = -6^{\frac{3}{5}} \ \ , \ \ \beta = 6^{\frac{1}{5}} \ ,
\eea
leads to
\bea
\label{e:PI_bis}
\frac{d^2 w}{dt^2} = 6 w^2 + t \ ,
\eea
namely the Painlev\'e-I equation in (\ref{e:Painleve-I}).

However, this is not enough to identify the specific  Painlev\'e-I transcendent,
relevant in this orbital transition to plunge setting. This involves choosing
a particular solution to (\ref{e:Ori_Thorne}) under the pertinent boundary conditions.
In this sense, Ori and Thorne look for the unique solution matching
smoothly the adiabatic inspiral, concluding that the correct asymptotics are
\bea
\label{e:asymptotics_PI}
X \sim \sqrt{-T} \ \ , \ \  T\to -\infty \ .
\eea
The solution should be regular (in particular having poles) for all $T<0$.
Remarkably this, together with the asymptotics (\ref{e:asymptotics_PI})
fixes the solution to Eq. (\ref{e:Ori_Thorne}): the relevant Painlev\'e-I
transcendent is the so-called Boutroux's {\em tritronqu\'e} solution.

In particular it was proved in \cite{JosKit01} that such  tritronqu\'e
has no poles for negative values of $T$, the first pole happening at
$t_0\sim 2.3841687$ with monotonically decreasing behavior in
$(-\infty, t_0)$~\footnote{The discussion in \cite{JosKit01}
deals with a version of  (\ref{e:PI_bis}), in the form $y'' = 6y^2 -x$,
so no poles are found for $x>0$ and the solution is monotonically decreasing in $(t_0, \infty)$.}.
As a non-trivial check of the consistency with the solution found
in \cite{Ori:2000zn}, compare the described behavior of the tritronqu\'e
solution with the one of the numerical solution in Fig. 3 of~\cite{Ori:2000zn},
in particular diverging at $T_{\mathrm{plunge}}$
\bea
T_{\mathrm{plunge}} = \beta t_0 =  6^{\frac{1}{5}}  t_0 \sim 6^{\frac{1}{5}}\cdot  2.3841687 \sim 3.41167 \ ,
\eea
(using Eqs. (\ref{e:PI_rescaling}) and(\ref{e:PI_rescaling_values}))
that refines the value in \cite{Ori:2000zn}, namely $T_{\mathrm{plunge}}\sim 3.412$.

In sum, we identify here that the relevant  Painlev\'e-I
transcendent in the Painlev\'e-I equation appearing in the
orbital transition to plunge at the ISCO~\cite{Ori:2000zn,Compere:2021iwh,Compere:2021zfj}
is Boutroux's tritronqu\'e solution, known to be associated with
universality properties~\cite{DubGraKle09} in the setting of 
critical behavior of Hamiltonian perturbations of nonlinear
hyperbolic PDEs~\cite{Dubrovin2006}. This offers a tantalizing avenue
to connect the ODE orbital EMRI problem with universality properties
of an underlying effective PDE dynamics describing finite mass-ratio binaries.

\subsection{EMRI quasi-circular orbits: Painlev\'e-II dynamics}
\label{e:EMRI-PII}
We explore now the adaptation of Rajeev's argument to a gravitational setting.
The idea in the present discussion is to make a first exploration of the EMRI problem
by generalizing Rajeev's by taking Eq.  (\ref{e:LL_eqs}) as the starting point. This
is a constraining condition and more systematic and general
studies exploraring systematically the Painlev\' e property of EMRI ODEs are needed,
our discussion here only representing a first step but already illustrating the strategy.

In this spirit, a natural gravitational counterpart of the Landau-Lifshitz equations
in the electrostotatic setting consists in  considering 
the leading order in the Post-Newtonian radiation-reaction equations for an EMRI binary system.
In the appropriate (harmonic) gauge,
the latter can be written (see e.g. \cite{Loutrel:2018ydu})
\bea
\label{e:a_RR_PN}
\frac{d\vec{v}}{dt} &=& -\frac{M}{r^2} + \vec{a}_{\mathrm{RR}} \nn \\
\vec{a}_{\mathrm{RR}} &=& \frac{8}{5}\eta \frac{M^2}{r^3}\left[\left(3v^2 + \frac{17}{3} \frac{M}{r}\right)\dot{r}\hat{r}
- \left(v^2+3\frac{M}{r}\right)\vec{v} \right] \ ,
\eea
where $\eta$ is the symmetric mass ratio and $\hat{r}=\vec{r}/r$ is the unit radial vector.

In the general orbital case, this cannot be written in the form (\ref{e:LL_eqs}). However, if we focus on the
quasi-circular case corresponding to the inspiral phase of the BBH problem, the approximations
$\displaystyle \dot{r}\sim 0,  v^2 \sim \frac{GM}{r}$ permit to rewrite Eq. (\ref{e:a_RR_PN}) in the form (\ref{e:LL_eqs})
with  $\displaystyle\tau= \frac{16}{5}\eta M$ the damping time scale and
\bea
\label{e:U_W_grav}
 U(r)= -\frac{M}{r} \quad , \quad 
W(r)= -\frac{M^2}{r^2}  \ ,
\eea
that is, $U(r)$ is the Keplerian potential, whereas $W(r)$ provides  a tidal potential.
If we insert $\alpha=-M$, $\beta=-M^2$, $n=1$, $m=2$ in Eqs. (\ref{e:ODE_v2})
we obtain the corresponding equations in the gravotaitonal case. However,
if we follow the steps performed in the electrostatic case, in this case one
does not get to a Painlev\' e -II equation as in Eq. (\ref{e:PII_Elec}). 
The responsible is the term $\Delta$  in the second equation in (\ref{e:ODE_v2}).
So, in this particular approximation to the EMRI problem, Painlev\' e-II is
not reated to generic orbits.

However, there is a
particularly interesting dynamical setting in which this term $\Delta$ can be neglected.
Indeed, looking at (\ref{e:Delta}), for quasi-circular orbits
 $\dot{r}\sim 0$ we have $\Delta\sim 0$ (note that it is crucial to have $m=2$ in the
potential $W(r)$ for this approximation not to depend on $r$, only 
in $\dot{r}$). Moreover, this is indeed
self-consistent with the quasi-circular assumption enforced to pass from Eqs. (\ref{e:a_RR_PN})
to Eq.  (\ref{e:LL_eqs}).  Further reaching studies will need to start from 
more generic forms of  PPN radiation reaction equations (in the same sense
that Comp\`ere and K\"uchler's work 
\cite{Compere:2021iwh,Compere:2021zfj} systematically extends that of Ori and Thorne~\cite{Ori:2000zn})
and (probably) look for the fulfilment 
of the Painlev\'e property. We leave this for future work.

Restricting ourselves in  to such quasi-circular dynamical scenarios in  (\ref{e:LL_eqs}),
with (\ref{e:U_W_grav}), we can write
\bea
\label{e:ODE_Grav}
\frac{dL}{dz} &=& \frac{2M\tau}{r^2}L \nn \\
\frac{dr}{dz} &= &\frac{r^3}{4M^3\tau}L^2 + \frac{r^2}{M}z + \frac{2M\tau}{r} \  ,
\eea
and, from this, we obtain
\bea
\label{e:aprox_PII_Grav}
\frac{d^2L}{dz^2} = -\frac{1}{M^2}L^3 -4\left(\frac{\tau}{r}\right)z L - \frac{4M^2}{r^2}\left(\frac{\tau}{r}\right)^2 L   \ .
\eea
There are two key differences with respect to the electrostatic case in (\ref{e:PII_Elec}). The
first one is that the coefficients depend on $r$, and therefore are not constants, as required
in the Painlev\'e-II equation. This however is a mild obstruction precisely in the quasi-circular
regime we are considering. Indeed, in that regime we can approximate $r\sim r_o$ during the appropriate
timescales dictated by the radiation-reaction process. The second one refers to the
last term in the equation, absent in the Painlev\' e equation. However, this term is a second-order
term in $\tau$, that we have neglected from the very beginning. Therefore in the
setting we are discussing, this term must also be neglected~\footnote{Note that if the approximation
  $r\sim r_o$ is adopted, this term can be reabsorbed into the second one by
  a ``shift'' redefinition of $z$.}. Under these approximations we get
\bea
\label{e:PII_Grav}
\frac{d^2L}{dz^2} = -\frac{1}{M^2}L^3 -4\left(\frac{\tau}{r_o}\right)z L    \ .
\eea
that, again, can be reduced to the Painlev\' e-II equation under the appropriate scalings.

This discussion partially generalizes Rajeev's argument to the gravitational case
and makes conceptual contact with the Painlev\'e-I result in 
\cite{Compere:2021iwh,Compere:2021zfj}. The most important outcome is the illustration of the
role of Painlev\' e transcendents in the EMRI gravitational binary problem.
In the asymptotic reasoning spirit in \cite{Batte01}, we claim that such structures
play also a structural role, though recast in another form more akin to a PDE discussion,  
in the full BBH problem.

\section{Binary black hole dynamics: an integrability ansatz}
\label{s:BBH_PDE_integrability}
In the previous section the Painlev\'e-II transcendent has
been identified, in the extreme mass ratio limit,
as a relevant underlying structure in BBH 
dynamics. Consequently, the Painlev\'e-II will have an imprint in
the BBH waveform, but its actual identification is done at the level
of BBH orbital dynamics. Specifically, from a methodological
perspective, the approach to BBHs above  separates: i) an orbital
non-linear system described (in the proper asymptotic limit)
by an integrable ODE, ii) the (linear) waveform emitted by such
system.

In this section we approach the problem from a PDE perspective.
Instead of considering the full Einstein equation system
---as done, e.g.,  in numerical BBH simulations---
we adopt an  asymptotic approach aiming at
identifying and focusing on the underlying mechanism(s).
Rather than starting from Einstein equations and 
proceeding in a `top-down' scheme, the effective approach
we adopt here is a `bottom-up' one in which, in particular,
we follow the methodology in the ODE treatment above,
separating the dynamics of the background and that of
wave propagating on it.

\subsection{Fast and slow degrees of freedom: linear waves over integrable dynamical backgrounds}
\label{s:fast_slow}
Observational evidence, supported by numerical simulations of the full non-linear Einstein
equations as well as the success of effective semi-analytic treatments well beyond their natural
regimes of application, strongly indicate that waveform dynamics in the BBH problem
displays ingredients with effectively linear character. On the other hand, the ODE treatment of the
previous section suggests a non-linear dynamics with integrability playing a structural role.
This suggests a methodological approach in which we  effectively separate dynamics
with different time scales, into ``fast'' and ``slow'' 
degrees of freedom. Specifically, we consider:

\begin{itemize}
\item[i)]  Fast degrees of freedom subject to linear dynamics, corresponding
  to the propagating  and observed waveform.
  
\item[ii)] Slow degrees of freedom subject to non-linear 
  (integrable) dynamics providing the background for waves.

\end{itemize}
We propose to address the effective BBH dynamics in terms of the coupling of such
separated fast and slow dynamics~\footnote{We notice that, on the one hand,
  such a separation of scales is the natural setting for multiscale
treatments to reduce full dynamics to asymptotic descriptions targeting
a particular phenomenon. On the other hand, it evokes the ``wave-mean flow'' approach
in fluid dynamics \cite{buhler_2014}. Concomitantly, the transfer of concepts and tools 
from (dispersive) fluid dynamics will be play a fundamental role in the following.}.

\subsubsection{A simplified one-dimensional effective toy model}
To fix ideas,  in the following we will focus on a highly
simplified model in which spatial dynamics are effectively one-dimensional~\footnote{Although
  the main motivation for adopting this toy model is the technical simplication of the
  equations, specifically in the discussion of integrability in terms
  of the `inverse scattering transform', the model can also be of physical relevance in the
  setting of the description of the BBH dynamics in a non-inertial frame
  rotating with the BHs. This would be a reminiscent of the strategy adopted
  when studying Lagrange points in the `restricted three body problem' in celestial
  mechanics (cf. e.g. \cite{MurDer99}).}. Denoting formally by $\phi$ the
fast linear degrees of freedom, and by $u$ the slow (background) degrees of freedom,
we consider a PDE dynamical system in which the fast dynamical
degrees of freedom  satisfy a linear wave equation, propagating on
an effective potential $V(x,t;u)$
\bea
\label{e:fast_degrees}
\left(\frac{\partial^2}{\partial t^2} - \frac{\partial^2}{\partial x^2} +V(x,t; u)\right)\phi = S(x,t; u) \ ,
\eea
where the effective potential $V$ is determined by the background slow degrees of freedom
$u$, that and satisfy a dynamical equation
\bea
\label{e:slow_degrees}
\partial_t u = F(u, u_x, u_{xx}, \ldots) \ .
\eea
Note that the modelling of the fast degrees of freedom include a
dynamical forcing $S$ in terms of a source possibly depending
of the slow degrees of freedoms. This will be crucial in the waveform model.

\subsubsection{A dispersive Ansatz on background integrable non-linear dynamics.}
\label{s:Dispersion_non-linear}
The role of the Airy function in the BBH waveform model in \cite{JarKri21}
suggests that the underlying mechanism behind the simplicity and universality
in BBH mergers includes a {\em dispersive} ingredient.
With this motivation, and also as methodological choice
to constraint anf guide the choice of possible models, we make the hypothesis that the background
dynamics is effectively (weakly) {\em dispersive}.

This is a strong assumption in the setting of general relativistic gravitational dynamics.
Such hypothesis can be motivated in an asymptotic ``coarse-grained''description
that averages over different background characteristic scales, the latter
inducing a qualitatively different scattering behaviour for waves of different lengthwave.
However, the ultimate validity of such dispersive hypothesis can be justified only
in terms of the obtained results.
As we will see below, this allows in particular to make explicit contact with the
Painlev\'e-II trascendent identified in the inspiral phase.

In particular, motivated by the identification of Painlev\'e-II at EMRI BBH dynamics,
and using the connection of
Painlev\'e transcendents with integrable systems,
we bestow a structuring role to the notion of integrability. 
It provides a concrete guiding hypothesis aiming at accounting for the observed simplicity
and universality in the BBH dynamic providing a methodological 
guideline unifying the treatment along the different regimes in the entire
BBH evolution, we grant a structuring role to the notion of integrability.
Specifically, in our fast-slow dynamical coupling scheme, we make the
hypothesis of integrability for the  non-linear dynamics for the slowly background,
over which the fast degrees propagate.

\subsection{Non-linear background dynamics: KdV-type models}
\label{s:KdV_models}
Once in a dispersive and non-linear setting, we make the following fundamental
remark  (cf. e.g. \cite{AblSeg81}):
{\em KdV-type equations provide ``universal'' models for weakly dispersive and weakly
  non-linear wave systems}.

Specifically, in the KdV equation introduced in (\ref{e:KdV})
\bea
\label{e:KdV_text}
\partial_t u + 6 u\partial_xu + \partial^3_{xxx}u = 0 \ ,
\eea
the advection term accounts for non-linearity, as in the Burgers' equation,
whereas the third derivative term corresponds to dispersion.
In spite of the crucial role of non-linearity, it is instructive
to start with the linearized case.

\subsubsection{Linear case: Airy}
\label{s:KdV_linear}
Let us consider the linearized KdV equation
\bea
\label{e:KdV_linear}
\partial_t u + \partial^3_{xxx}u = 0 \ ,
\eea
and its solutions satisfying $u\to 0$ when $|x|\to \infty$, with initial data
$u_0(x) = u(t=0,x)$ satisfying (note we only consider here real solutions)
\bea
\label{s:u_0_normalization}
\int_{-\infty}^\infty (u_0(x))^2 dx < 0 \ .
\eea
Specifically, following~\cite{Ablow11} (see also \cite{AblSeg81}),
we consider its asymptotic solutions corresponding to the
regions: i) $x/t<0$, ii) $x/t>0$, and iii) $x/t \to 0$.

First we write the (formal) solution of (\ref{e:KdV_linear})
in terms of the Fourier transform $\hat{u}_0(k)$ of $u_0(x)$
\bea
u(x,t) = \frac{1}{2\pi}\int \hat{u}_0 e^{i(kx -\omega(k)t)}dx = \frac{1}{2\pi}\int \hat{u}_0 e^{i(kx + k^3t)}dx \ ,
\eea
where the employed relation dispersion $\omega(k)=-k^3$, followed
by inserting the first expression into (\ref{e:KdV_linear}).

\paragraph{Region $x/t<0$.} Here a {\em stationary phase} method is employed.
Writing
\bea
\label{e:u_Fourier}
 u(t,x)=\frac{1}{2\pi}\int \hat{u}_0 e^{i\phi(k)t)}dx \ ,
 \eea
 with $\phi(k) = k(x/t) + k^3$, for given $x$ and $t$, we determine
 the stationary points $k_0$ from the vanishing of $\phi'(k)$, that is
 \bea
 \label{e:stationary points}
 \phi'(k_0) &=& \frac{x}{t} + 3 (k_0)^2 = 0 \nn \\
 (k_0)_{\pm} &=& \pm \sqrt{\frac{-x}{3t}} 
 = \pm \sqrt{\left|\frac{x}{3t}\right|} \ .
 \eea
 Using then $\phi''(k)$, expanding to the second
 order, and summing over the stationary points in the
 standard form of the stationary phase expansion,
 we find 
 \bea
 u(t,x) &\sim& \frac{\hat{u}_0((k_0)_+)}{\sqrt{2\pi t|\phi''((k_0)_+)|}}
 e^{i\phi((k_0)_+)t+i\mu_+ \pi/4} \nn \\
 &+& \frac{\hat{u}_0((k_0)_-)}{\sqrt{2\pi t|\phi''((k_0)_-)|}}
 e^{i\phi((k_0)_-)t+i\mu_- \pi/4} \ ,
 \eea
 with $\mu\pm = \mathrm{sgn}(\phi''((k_0)_\pm)=\pm 1$. Inserting the
 expressions for $\phi((k_0)_\pm)$ and $\phi''((k_0)_\pm)$ and
 writing $\hat{u}_0(k)=|\hat{u}_0(k)|e^{i\hat{\psi}_0(k)}$,
 we get
 \bea
 \label{e:linear_KdV_cos}
 u(t,x)\sim \frac{\left|\hat{u}_0\left(\sqrt{\left|\frac{x}{3t}\right|}\right)\right|}{\sqrt{\pi t}
   \left|\frac{3x}{t}\right|^{1/4}}\cos\left(2\left|\frac{x}{3t}\right|^{3/2}t  
  -\frac{\pi}{4} -\hat{\psi}_0\left(\sqrt{\left|\frac{x}{3t}\right|}\right)\right)
\eea

\paragraph{Region $x/t>0$.} In this regime a related asymptotic method, namely
the {\em steepest descent}, is better adapted to extract the dominant
asymptotic behaviour. The solution is now written as a comple integral 
\bea
 u(t,x)=\frac{1}{2\pi}\int_C \hat{u}_0 e^{t\phi(k))}dk \ ,
\eea
where $\phi(k)$ is complex-valued. After finding the saddle points satisfying
$\phi'(k_0)=0$, the contour $C$, initially $]-\infty, \infty[$,
    is then deformed in the complex plane  using Cauchy's theorem so that $\mathrm{Im}(k)$ is constant
    and given by $\mathrm{Im}(k_0)$. This defines the steepest descent path $C'$.
    Implementing this scheme (see details in \cite{Ablow11}) and choosing the only compatible
    saddle point, one finally gets
    \bea
    \label{e:linear_KdV_exp}
    \displaystyle
    u(t,x) \sim \frac{\hat{u}_0\left(i\sqrt{\left.\frac{x}{3t}\right.}\right)}{2\sqrt{\pi t}
   \left(\frac{3x}{t}\right)^{1/4}}e^{-2\left(\frac{x}{3t}\right)^{3/2}t}
    \eea

    \paragraph{Transient region $x/t\to0$.} Looking at Eqs. (\ref{e:linear_KdV_cos}) and (\ref{e:linear_KdV_exp})
    in the respective $x/t<0$ and $x/t>0$ regions, we notice the particular combination of $x$ and $t$ entering
    in the  asymptotic expressions of $u(t,x)$. This suggests to look for a so-called {\em self-similar
      solution} at the transient regime, namely an Ansatz for the functional form o $u(t,x)$ as
    \bea
    \label{e:similarity}
    u(t,x) \sim \frac{1}{(3t)^p} f\left(\frac{x}{(3t)^q}\right) \ ,
    \eea
    Injecting (\ref{e:similarity}) into (\ref{e:KdV_linear}) we get an ODE for $f$
    in its natural variable $\eta=x/(3t)^{q}$. Imposing that the ODE does not depend on $t$
    fixes $q=1/3$. In this linear setting $p$ is not fixed at this level
    and we must resort to the connection with the other asymptotic regimes above
    to render $p=1/3$. As a result, one gets the Airy equation as the ODE satisfied by the
    function $f$
    \bea
    \label{e:Airy_eta}
    \frac{df}{d\eta^2} - \eta f = 0 \ .
    \eea
     This permits to write the solution to (\ref{e:KdV_linear})
     in the transient region $x/t\to 0$ in terms of the Airy function, namely
     \bea
     \label{e:linear_KdV_xt0_v1}
     u(t,x) \sim \frac{1}{(3t)^{1/3}} \mathrm{Ai}\left(\frac{x}{(3t)^{1/3}}\right) \ .
     \eea
     This expression provides a first contact with universality, with the remarkable
     fact of involving a structural connection 
     with the notion of self-similiarity in our PDE context.
     In particular, the Airy function pattern provides a universal profile of the transient
    at $x/t\to 0$ between the oscillatory behaviour at $x/t<0$ and the damped one
    at $x/>0$. The Airy function is the universal special function
    playing,  for linear 'turning point' problems,
    the analogous role played by the $\tanh(x)$ function for
    shocks, through Eq. (\ref{e:universal_shock}). That is, paraphrasing
    the statement for Burgers' equation in this context: {\em all linear turning points transitions
    look the same}.
    
     A more systematic (an improved)  manner of obtaining an asymptotic solution in this regime
    makes use of the integral representation of the Airy function
    \bea
    \mathrm{Ai}(\eta) = \frac{1}{2\pi}\int_{-\infty}^\infty e^{i(s\eta + s^3/2)}ds \ .
    \eea
    Then, starting from the Fourier representation of the solution
    (\ref{e:u_Fourier}), making the variable change $k=s/(3t)^{1/3}$
    and expanding $\hat{u}_0(k)=\hat{u}_0(s/(3t)^{1/3})$ 
    to the first order in $s$ in Taylor series around $s=0$, one can finally
    write    
    \bea
    \label{e:linear_KdV_xt0}
    u(t,x) \sim \frac{\hat{u}_0(0)\mathrm{Ai}(\eta)}{(3t)^{1/3}}
    &&-  \frac{i\hat{u}'_0(0)\mathrm{Ai'}(\eta)}{(3t)^{2/3}} \ ,
    \eea
    where both terms are indeed self-similar solutions of the linear KdV equation.

    \paragraph{Uniform asymptotic expansion.} Once the asymptotic behaviours in each $x/t$ region have been identified,
    a {\em matched asymptotic expansion} can be employed to join the different
    regimes. This actually permits to fix $p=1/3$
    in Eq. (\ref{e:similarity}), in this discussion of linear (\ref{e:KdV_linear}).
    Remarkably, using the asymptotic expressions of the Airy function
    one can indeed write an asymptotic expression valid in the three regimes,
    namely
    \bea
    \label{e:u_linear_KdV_global}
    u(t,x) \sim  &&\frac{\hat{u}_0(k_0)+\hat{u}_0(-k_0)}{2}\frac{1}{(3t)^{1/3}} \mathrm{Ai}\left(\frac{x}{(3t)^{1/3}}\right)
     \\       
     && -  \frac{\hat{u}_0(k_0)-\hat{u}_0(-k_0)}{2ik_0}\frac{1}{(3t)^{2/3}}\mathrm{Ai'}\left(\frac{x}{(3t)^{1/3}}\right)
     \nn \ ,
    \eea
    with $k_0=\sqrt{-x/(3t)}$.     
    The global expression (\ref{e:u_linear_KdV_global})
    represents a {\em slowly variating similarity solutions}, namely
    a linear combination of self-similar solutions (with $f$ given,
    respectively, by $\mathrm{Ai}$ and $\mathrm{Ai}'$)
    modulated by slowly varying factors 
    determined by the initial data, through its Fourier transform $\hat{u}_0(k)$.

    This expression is structurally similar to the one we have found for the
    diffraction patterns on a caustic in \cite{JarKri21}. This is not an accident, actually
    a {\em uniform asymptotic expansion}~\cite{CheFriUrs57,Berry76,BerUps80,KraOrl12} 
    can be employed for deriving expression (\ref{e:u_linear_KdV_global}), following exactly the
    same steps implemented in determining the universal patterns for caustics (cf. \cite{JarKri21}).
    Such uniform asymptotic expansion remains valid in the regimes in which (real or complex) ``critical points''
    $(k_0)_\pm=\pm\sqrt{-x/(3t)}$
    are separated and also when they degenerate at the ``caustic''. In particular, in our present
    problem this identifies a {\em (fold) catastrophe} in our PDE problem at the transient regime
    $x/t\to 0$.

    On the other hand, the physical role of this solution in terms of the
    Airy function and its derivative is very different to the one discussed
    in \cite{JarKri21}: whereas in the caustic discussion in \cite{JarKri21} the Airy
    solution accounts for the (linearly) propagating wave field (fast degrees
    of freedom) according detected at the observer emplacement,
    Eq. (\ref{e:u_linear_KdV_global}) describes (in the linearised approximation of Eq.
    (\ref{s:KdV_linear}))
    the dynamics of the slow degrees providing the ``effective potential''
    on which the fast degrees of freedom are scattered. 
    We develop further this point below, in the discussion of non-linear turning points
    in terms of the Painlev\'e-II trascendent.

\subsubsection{Non-linear case: Painlev\'e-II}
\label{s:KdV_nonlinear}
After considering the linear case (\ref{e:KdV_linear}) of the KdV equation
(\ref{e:KdV_text}), we move to the actual non-linear case.
From the linear case, we highlight three structural elements:

\begin{itemize}
\item[i)] {\em From Fourier analysis to the inverse scattering transform (IST)}.
  In the linear setting, the starting point for the
  study of the asymptotics of $u(t,x)$ in the different dynamical regimes is
  the Fourier expression (\ref{e:u_Fourier}) in terms of initial data.
  However, in contrast with the linear case, no use of the Fourier
  transform can be made in the non-linear setting. Remarkably,
  the IST scheme, provides a solution to the problem that can indeed
  be dubbed as a {\em non-linear Fourier transform}.
  Most importantly at a structural level, such IST scheme actually characterizes 
  what we mean by integrability at the PDE level in terms of the
  so-called {\em Painlev\'e test} .

\item[ii)] {\em Self-similar solution and universality}.
  The transient asymptotics of the solution are captured by a self-similar
  solution, that implements universality through a universal
  pattern. In this spirit, we will enforce the solution $u(x,t)$
  to have a self-similar character.

\item[iii)] {\em Matching conditions and uniform asymptotics}. The self-similar
  solution should provide also the key to match asymptotics corresponding
  to oscillation and damped behaviours, as the
  Airy function does in the linear case. That is, the self-similar
  solution should implement a non-linear 'turning point' matching
  providing a uniform asymptotic description through the merger.
  
\end{itemize}
In particular, the combination of the second and third points suggests to focus, rather
than on the standard non-linear KdV equation (\ref{e:KdV_text}),
on the modified one  m-KdV in Eq. (\ref{e:modified_KdV}),  namely 
\bea
\label{e:modified_KdV_text}
u_t - 6u^2\partial_xu + \partial^3_{xxx}u = 0 \ .
\eea
This is a KdV-like equation sharing the dispersive properties of
KdV and whose solutions are related to those of KdV by a (Miura/)B\"acklund
Ricatti transformation (that do not introducing new movable critical points
\cite{AblSeg77}). In our setting, the m-KdV has the virtue of providing a
direct connection between the PDE dynamics of the background non-linear
degrees of freedom, on the one hand, 
and the  Painlev\'e-II transcedent identified in the BBH ODE dynamics, on the
other hand. And this is done precisely by means of enforcing a
self-similar solution to (\ref{e:modified_KdV_text}) that, as discussed
above in the linear case, is intimately related to the realization of universality.

Specifically, focusing on the transient asymptotics $x/t\to0$ and
inserting the self-similar Anstaz (\ref{e:similarity}) into
Eq. (\ref{e:modified_KdV_text})
and imposing the ODE to be time independent
fixes {\em both} $p$ and $q$ to $p=q=1/3$ (note the contrast with
the linear case, where only $p$ was fixed, requiring the full
asymptotic matching to fix $q$). 
The resulting ODE, analogous to Eq. (\ref{e:Airy_eta}), is then
\bea
\label{e:Painleve-II_alpha=0}
\frac{d^2 f}{d\eta^2} -\eta f -2 f^3  = 0 \ ,
\eea
namely the Painlev\'e-II equation in (\ref{e:Painleve_equations}) with $\alpha=0$.
This leads   \cite{AblSeg77b,SegAbl81} to the non-linear analogue of solution (\ref{e:linear_KdV_xt0_v1})
\bea
u(t,x) \sim \frac{1}{(3t)^{1/3}}\mathrm{P_{II}}\left(\frac{x}{(3t)^{1/3}}\right)  \ ,
\eea
in the asymptotic $x/t\to0$ merger regime. In other words, the Painlev\'e-II
transcendent provides the special function fixing the
universal pattern of non-linear turning points.

An analogue of the more general Eq.~(\ref{e:linear_KdV_xt0}) in the linear case,
matching the uniform asymptotic expansion (\ref{e:u_linear_KdV_global}),
can also be derived ~\cite{SegAbl81} in the present non-linear setting~\footnote{Notice that
  Eq. (\ref{e:P_plus_P'}) is much more rigid that the corresponding solution in terms
of the Airy function for the linear equation and the free-data
$\hat{u}(k_0)$ in (\ref{e:u_linear_KdV_global}), since the non-linearity
of m-KdV does not permit to adjust the solution (\ref{e:P_plus_P'})
by making linear combinations of solutions
with free  slowly varying modulation factors. An approach to introduce
such a freedom for a slowly varying modulation would be 
one needs to introduce this at the level of the equations,
suggesting a connection with so-called Whitham equations.
More generally, IST is used for solving in terms of initial data.}
\bea
\label{e:P_plus_P'}
u(t,x) \sim
\frac{1}{(3t)^{1/3}} \mathrm{P_{II}}\left(\frac{x}{(3t)^{1/3}}\right)
     +  \frac{1}{(3t)^{2/3}}(\mathrm{P_{II}})'\left(\frac{x}{(3t)^{1/3}}\right) \ ,
     \eea
     that, given the asymptotics described in section \ref{s:P_II_turning_point_model},
     directly generalises  the Airy-like solution (\ref{e:linear_KdV_xt0}) to the
     non-linear case.

     As anticipated at the end of section \ref{s:BBH_PDE_integrability}, formal contact
     has been made with the Painlev\'e-II structures found in  the orbital dynamics of
     damped binary EMRI dynamics discussed in section~\ref{s:Painleve_ODE}.
     But, as in the ODE orbital case ---and as discussion after Eq. (\ref{e:u_linear_KdV_global}))---
     this refers to the `slow degrees of freedom'.  Therefore, the dynamics determined by
     (\ref{e:P_plus_P'}) does not directly describe to the
     BBH waveform, but rather the background dynamics generating such waveform.
     Indeed, in order to make contact with the waveform dynamics and
     connect with the model presented in \cite{JarKri21}, a further step is needed.

\subsection{Heuristics on the slow-fast BBH dynamics coupling: $\mathrm{P_{II}}$-driven wave equation}
\label{s:heurisitics}
In this section we discuss some heuristic elements meant as a basis for the
development of an approximate (asymptotic~\cite{Batte01}) scheme for the PDE
treatment of the BBH problem. We organise the discussion in term
of the following points
\begin{itemize}
\item[i)] {\em Dual-frame approach: irrotational fast dynamics over corrotating slow dynamics}.
  As discussed above, slow degrees of freedom $u$ serve as a background over which fast degrees
  of freedom propagate. During the inspiral phase the inner wave zone presents an approximately
  stationary behaviour~\cite{WheRom99,WheKriPri00,WheBeeLan02}. In this setting, a corrotating
    frame seems adapted for describing the slow dynamics~\footnote{This is a reminiscent of the
      corrotating coordinate system employed in the ``restriced three-body'' problem in celestial
      mechanics, in particular in the discussion of the ``Lagrange points''.}. On the other
    hand, the fast degrees of freedom are more
    naturally described in a coordinate system that is irrotational with respect to
    the far wave zone. In other words, coordinates in (\ref{e:fast_degrees}) must be understood
    in an irrotational frame, whereas coordinated in (\ref{e:slow_degrees}) should be understood
    rather in a corrotating system.

  \item[ii)] {\em Integrability, self-similar solutions, universality and IST for background dynamics:
    the mKdV model}. The dynamics
    describing the slow degrees of freedom are modelled by a universal self-similar solution
    to an integrable dispersive non-linear PDE. In particular, such a self-similar solution
    is structurally related to a Painlev\'e transcendent, specifically to
    $\mathrm{P_{II}}$ in our BBH setting. Such reducibility of the non-linear PDE
    to a  Painlev\'e equation is intimately related, through the so-called Painlev\'e test,
    to the integrability of the PDE in the sense of being solvable by IST.
    This provides, in particular, an infinite set of conserved quantities that constraint
    and simplify the dynamics. A simple model for this scheme is provided by the
    mKdV equation (\ref{e:modified_KdV}) whose universal self-similar solution is
    given the expression (\ref{e:P_plus_P'}) in terms of  $\mathrm{P_{II}}$ and its derivative.
    This is the PDE counterpart of the ODE $\mathrm{P_{II}}$ terms in the EMRI BBH
    problem of section \ref{s:Painleve_ODE} for the orbital (background) dynamics.
    The imprint of $\mathrm{P_{II}}$ into the waveform is realised through the coupling to
    the fast degrees of freedom.

  \item[iii)] {\em $\mathrm{P_{II}}$-pattern in inspiral and merger waveform: background dynamics
    as a forcing term}. The key mechanism to imprint the Painlev\'e-II-like
    pattern into the waveform, therefore providing a $\mathrm{P_{II}}$ generalization
    of the Airy model in \cite{JarKri21}, is the forcing by the source term $S(x,t; u)$
    in Eq. (\ref{e:fast_degrees}). Indeed, ignoring initial transients, 
    such linear forced system is controlled by the driving terms $S$.
    Such driving term is determined by two ingredients: a) the background solution
    $u(t,x)$, namely shaped by the $\mathrm{P_{II}}$ transcendent, b) a modulation
    given by the rotation of the (slow-dynamics) corrotating frame with respect to the
    (fat-dynamics) irrotationnal one. In particular, in an approximation in which
    we neglect the ``potential term'' $V(x;u)$ in the homogeneous (left hand side)
    part of (\ref{e:fast_degrees}), we can write~\footnote{Such an expression
      is valid for odd spatial dimensions, in particular the dimensions one here discussed
    and the physical dimension three, with appropriate spatial decaying factors.}
    \bea
    \label{e:Green_function}
    &&\phi(t,x) = \int G(t,x;t',x') S(x',t'; u(t',x'))  \\
    && \hbox{where} \quad G(t,x;t',x') \sim \delta\big(t'-(t-|x-x'|)\big) \nn \ .
    \eea    
    The interest of such an expression is that the waveform is now described by $\phi$,
    that captures the pattern in the forcing term and transport it along the characteristics.
    When including back the $V(x;u)$ term in (\ref{e:fast_degrees}), resonant phenomena can occur
    when the source $S(x,t; u)$  triggers the resonant frequencies associated with the
    potential~\footnote{If such a resonant mechanism can play a role in the merger
    transition, specifically in the properties of the maximum peak, 
    is a question that requires specifically devoted study.}.
    In summary, the inspiral and merger transient waveform is controlled by a forcing
    mechanism driven by the background integrable dynamics, 
    leading to a ``modulated-$\mathrm{P_{II}}$'' model for the waveform that makes
    contact and extend the Airy-waveform model in \cite{JarKri21}.

  \item[iv)] {\em Merger transition to ringdown: waveform dynamics from background potential}.
    The separation between fast and
    slow degrees of freedom makes only sense as long as the associated timescales
    are not commensurate. This is no longer the case in the merger. At this point
    a ``phase transition'' occurs that, as discussed in \cite{JarKri21}, does not
    happen at the peak of the merger waveform, but a bit later associated with
    the $t=0$ value in the $\mathrm{P_{II}}$ or Airy ODE, respectively Eqs. (\ref{e:Painleve-II_complete})
    and (\ref{e:Airy_equation}). At this point the
    slow background dynamics freezes, the forcing term $S(x,t; u)$
    ``decouples'' (actually disappears) and the waveform dynamics is determined
    by the potential $V(x; u)$,  so the late ringdown is controlled by the
    resonant properties of $V(x; u)$. The latter
    is constructed from a time-independent (solitonic) solution
    of the integrable background dynamics, so integrability underlies the associated scattering theory.
\end{itemize}

\section{Integrability: a unifying principle for background BBH non-linear dynamics.}
\label{s:integrability_background_dynamics}
Integrability is proposed in the previous sections as a structural guideline to
account for the simplicity and universality features in the BBH merger waveform
and, more generally, BBH dynamics.
Such a proposal is supported by the finding of the role played by
Painlev\'e transcendents in the (ODE) binary dynamics in the inspiral phase
and, on the other hand, by the integrability features  present in the
description of the BH scattering in the late merged BH, revealing
the sound link to KdV structures responsibke of features
such as BH quasi-normal mode
isospectrality~\cite{Chandrasekhar:579245,Glampedakis:2017rar,Lenzi:2021njy,Lenzi:2021wpc}.
Between these two phases, in the previous section we have proposed
that integrability indeed extends through the merger, discussing a toy-model
built on a fast-slow separation of gravitational degrees
in which the slow non-linear dynamics would be integrable~\footnote{
  Instances of such a paradigm involving the linear propagation of perturbations on a
  background solution to non-linear integral equations
  have been explored in other physical settings
  as, e.g. in the propagation of (linear) perturbations on the
  background of topological-soliton skyrmions in the context of  nuclear
  physics (we thank M. Semenov-Tian-Shansky for bringing our attention
  to this type of models).}, with
self-similarity and dispersion as key ingredients to account for universality
and connect woth Painlev\'e transcendents.
In this section, we aim at abstracting some of the relevant structural elements
that would permit to go beyond the particular (ad hoc) KdV-like models discussed above.

\subsection{Integrability and inverse scattering transform: a `bottom-up'
  approach to BBHs}
Ideally, one would like to start from Einstein equations and, by an
appropriate reduction process~\footnote{The use of asymptotic tools~\cite{bender1999advanced,Ablow11}
  permits partial implementations of this program in the fluid dynamics context,
  namely in the setting of dispersive non-linear hydrodynamics.}, arrive to an integrable PDE system
approximating (background) BBH dynamics.
Although performing such a ``top-down'' approach would be the ultimate objective,
here we focus on a (more humble) ``bottom-up'' approach, limiting ourselves
to indicate some elements to explore in the ``top-down'' setting.

\subsubsection{Integrability, Inverse Scattering Transform and Lax pairs}
\label{e:integrability}
Although there is no universal definition of integrability for
(non-linear) PDEs, a standard characterisation relies on the possibility
of transforming the non-linear PDE into an equivalent linear problem,
namely the so-called {\em inverse scattering transform} (IST).
The key ingredient of this method is the possibility of writing
the non-linear equation (\ref{e:slow_degrees}) in a Lax representation,
namely finding operators $L$ ad $A$, depending on  $u$, $u_x$, $u_{xx}$...,
such that  (\ref{e:slow_degrees}) is equivalent to the equation
\bea
\label{e:Lax_pair}
L_t = [L, A] \ , 
\eea
In general, the operators $L$ and $A$ depend on a spectral
parameter $\lambda$~\cite{sokolov2020algebraic}. We focus here on the case that this
parameter $L$ appear as an eigenvalue of $L$, that is
\bea
\label{e:eigenvalue_L}
L \psi = \lambda \psi \ .
\eea
If we write the evolution of the eigenfunctions as
\bea
\label{e:eigenfunction_evolution}
\psi_t = A \psi \ ,
\eea
then Eq. (\ref{e:Lax_pair}) is equivalent to the time independence of $\lambda$
\bea
\label{e:isospectrality}
\dot{\lambda} = 0
\eea
In other words, equation (\ref{e:slow_degrees})
(or its Lax pair representation (\ref{e:Lax_pair})),
can be seen as compatibility condition between
the two linear operators in (\ref{e:eigenvalue_L}) and (\ref{e:eigenfunction_evolution}).

The crucial consequence of the Lax pair rewriting (\ref{e:Lax_pair},
is that it permits recast the resolution of the initial value
problem of  (\ref{e:slow_degrees}) in terms of the direct and
inverse scattering problems of the operator $L$, understood
as the operator describing the (linear) scattering of a wavefunction
$\psi$. In the realisation in the KdV case, $L$ is explicitly the one-dimensional Schr\"odinger
operator (see e.g.~\cite{Dunajski:2010zz})
\bea
\label{e:L_KdV}
L = -\frac{d^2}{dx^2} + u(x,t) \ .
\eea
In particular, on the one hand, the direct scattering problem permits to determine
scattering data, that we formally denote as  $\{a(\lambda), b(\lambda), \ldots\}$,
in terms of the scattering operator $L$ (namely the potential $u$ in the KdV case,
at at given fixed time) when appropriate boundary conditions are imposed on $\psi$.
On the hand, the inverse scattering problem permits to retrieve
$L$ (i.e. $u(x, t=t_o)$) from scattering data $\{a(\lambda), b(\lambda), \ldots\}$,
namely using the Gelf'and-Levitan-Marchenko (GLM) equations \cite{Gelfa51,March55}.
Using these elements, the inverse scattering transform (IST) (cf. \cite{GarGreKru67,AblSeg81,KleSau22}
proceeds according to the follwing scheme:
\begin{itemize}
\item[i)] {\em Direct scattering: from initial data $u(x, t=0)$ to scattering data}. From the initial data $u(x, t=0)$ of
  (\ref{e:slow_degrees}) we determine the operator $L(t=0)$ 
  at time $t=0$. For instance, in the KdV case (\ref{e:L_KdV}) this fixes the potential.
  From $L(0)$ one then determines the scattering data at $t=0$: $\{a(\lambda, t=0), b(\lambda, t=0), \ldots\}$.

\item[ii)] {\em Evolution of scattering data}. Using equation the isospectral time evolution Eq. (\ref{e:Lax_pair})
  one then determines the evolution $\{a(\lambda, t), b(\lambda, t), \ldots\}$ of the scattering data.
  The isospectrality property (\ref{e:isospectrality}) underlies the existence
  of an infinite number of conserved quantities, a key feature of the integrability.

\item[iii)] {\em Inverse scattering transform: from scattering data to solution $u(x, t)$}.
  From the evolved scattering data,
  the GLM equations of inverse scattering permit to retrieve $L(t)$, namely the
  evolved potential $u(x,t)$ in the KdV case. This involves the resolution of a linear
  integral equation (namely a Riemann-Hilbert boundary problem).

\end{itemize}
This scheme can be assimilated to a non-linear equivalent of the Fourier
approach to solve a linear PDE: indeed steps i) and ii) above would correspond
to the calculation of the Fourier transform of the initial data, whereas step ii)
would correspond to the inverse Fourier transform to obtain the evolved solution.

\subsubsection{Painlev\'e test and self-similar solutions}
\label{e:Painleve_test}
How can we determine a priori if a equation  (\ref{e:slow_degrees}) is integrable?
Although there is no full algorithm to address this question for PDEs, the so-called
{\em Painlev\'e test} \cite{AblRamSeg80a,AblRamSeg80b,AblCla91} indicates that
a PDE that is reduced through a similarity transformation to a Painlev\'e ODE
(more generally, an ODE satisfying the {\em Painlev\'e property} discussed in section
\ref{s:Painleve_transcendents}) is probably solvable through a IST.
The test provides actually ony necessary conditions~\cite{Dunajski:2010zz}.

This remark is particularly important in our present case of BBH dynamics.
We remind that the main outcome in section \ref{s:Painleve_ODE} was the
identification of Painlev\'e transcendents (namely $\mathrm{P_{I}}$ and
$\mathrm{P_{II}}$) as relevant ODEs for the description of some  key
structural features of EMRI BBH orbital dynamics. The effort in section~\ref{s:BBH_PDE_integrability}
was to find a link with such  Painlev\'e transcendents from a PDE perspective,
this leading us to self-similar solution of certain dispersive non-linear PDEs.
In this BBH setting, it is natural to look at the Painlev\'e test from the
opposite perspective (justified, since the Painlev\'e test is not a sufficient
condition, but rather a necessary one): an integrable PDE in the sense of
being solvable by IST will admit a reduction to an ODE with the Painlev\'e
property, possibly a Painlev\'e transcendent.

Our  BBH interest on reductions to Painlev\'e ODEs justifies our focus on PDEs
integrable by IST. This is in particular the case of the toy-model in section
\ref{s:KdV_linear}, namely 
the modified-Korteweg-de Vries in Eq. (\ref{e:modified_KdV_text}), indeed IST
integrable and leading to $\mathrm{P_{II}}$~\cite{AblSeg77}.
But, in addition, the associated self-similar solutions to the correponding
PDEs have an interest on their own in our quest for universal features,
since their simple behaviour under appropriate re-scalings is a feature pointing
to the kind of simplicity and universality akin to the one happening critical phenomena~\cite{barenblatt_1996}.

The Painlev\'e test algorithm can be summarised~\cite{AblCla91,ConMus08,Dunajski:2010zz} as:
\begin{itemize}
\item[i)] Find the (Lie-)point symmetries of the studied PDE (this fixes the
  similarity transformations).
\item[ii)] Construct the associated ODE to the group-invariant solutions (self-similar
  solutions).
\item[iii)] Check the {\em Painlev\'e property} in the resulting ODEs asociated
  to self-similar solutions.
\end{itemize}

\subsection{Towards a top-down integrability approach to BBHs}
\label{s:top-down}
Understanding and characterising (aspects of) gravity at a fundamental
level in terms of integrable systems is a
challenging problem that has been addressed from different avenues
(see e.g. \cite{Frolov:2017kze} for a review).
In spite of the lack
of comprehensive theory, partial results revealing the important
role of integrability in gravity are available. We comment on
some aspects of potential relevance in the present context.

\subsubsection{Darboux covariance in the direct scattering problem}
\label{e:Darboux_cov}
Linearised Einstein equations in the context of black hole scattering theory
provide a first neat connection with integrability. Specifically, 
as revealed by Chandrasekhar~\cite{Chandrasekhar:579245},
odd and even-parity effective potentials in the stationary spherically
symmetric case (Schwarzschild and Reissner-Nordstr\"om) present  
the same transmission and reflection coefficients and
are isospectral in the associated quasi-normal mode frequencies.
These spectral features  can be
understood in terms of conserved quantities of the KdV equation,
that provides an isospectrality condition to be satisfied by the respective
effective potentials. A related but alternative approach to
this isospectrality in terms of Darboux transformations is presented
\cite{Glampedakis:2017rar,Yurov:2018ynn} (see also~\cite{Anderson:1991kx} for
an approach in terms of ``intertwining operators'').
However, it is in the recent work~\cite{Lenzi:2021wpc,Lenzi:2021njy}
where the connection to integrability, specifically through inverse scattering theory,
becomes apparent, as well as shedding light on the link
between the KdV equation and Darboux transformations (see also
\cite{Charalambous:2021kcz,BenAchour:2022uqo,Katagiri:2022vyz,hui2022ladder,Hui:2022vbh,Charalambous:2022rre} for other
hints on integrability and hidden symmetries in this perturbative setting).
Specifically, in \cite{Lenzi:2021njy} an infinite branch of new admissible
of odd/even effective potentials (with their associated master functions)
is identified, all related by Darboux transformations preserving the
spectral properties, leading to the notion of ``Darboux covariance''.
On the other hand, the role of the KdV equation as defining a flow
in the Darboux-related effective potentials is elucidated, making
full use of the Lax pair representation of the KdV equation and,
eventually, extending the universality of the infinite KdV hierarchy of conserved
quantities to the (infinite) branch of Darboux-related potentials.
This work is a most important one in our present context,
in particular regarding the integrability properties of the
effective potentials in the equation (\ref{e:fast_degrees})
satisfied by the linear degrees of freedom.

\subsubsection{Symmetry reductions of Einstein equations and integrability}
Staying at the level of the full non-linear Einstein equations (in contrast
with the linearised version discussed above), the enforcement of
symmetries provides another avenue towards integrability.
The archetype of this is the Ernst equation, that can indeed be solved through
an inverse scattering method~\cite{klein2005ernst}. In our present
BBH setting, let us comment on two avenues exploiting this symmetry reduction
procedure and, finally, a third avenue making a link such symmetry reductions and
self-dual Yang-Mills equations. The latter leads to the called Ward's conjecture,
that could provide a characterization of complete integrability in term of
twistor spaces potentially richer and beyond the scope of the inverse scattering
method. Specifically:
\begin{itemize}

\item[i)] {\em Ernst equation}.  Einstein equations under a stationary 
  and axisymmetric assumption can be written as an non-linear elliptic
  equation on a complex potential ${\cal E}$. Specifically, applying
  a projection formalism due to Ehlers \cite{Ehler57} and Geroch \cite{Geroc71}, the
  potential ${\cal E}$ lives
  in the two-dimensional space obtained by taking the quotient along
  the orbits of the two exisiting Killing vector fields.
    The resulting equation on ${\cal E}$  is the so-called  Ernst equation, that turns out to be completely integrable~\footnote{This
    is the so-called {\em elliptic} Ernst equation, obtained under imposing
    a timelike and and a spacelike Killing vector. If two spacelike Killing
    vector are enforced instead, one recovers the so-called
  {\em hyperbolic} Ernst equation, relevant in the study of colliding plane
  gravitational waves.}. Specifically, such equation can
  be reduced to a linear system through the inverse scattering method,
  presenting a Lax pair representation and the 
 Painlev\'e property discussed in section \ref{e:integrability}.

 Such integrability has been extraordinarily useful in the construction of
 stationary axisymmetric solutions in astrophysical settings(e.g.~\cite{Neuge79,KraNeu84,NeuKra69}).
 This property is intimately linked to the 
  existence of an infinite-dimensional group generated by the two commuting
  Killing vectors and acting as a symmetry on the space of solutions, namely the Geroch group.
  Interestingly, these notions have been explored in the BBH context,
  specifically in the study of stationary binary inspirals~\cite{WheRom99,WheKriPri00,WheBeeLan02}.
  As commented in section \ref{s:heurisitics} when sketching the {\em dual-frame approach},
  these ideas could be of much interest in the
  construction of integrable effective non-linear dynamics of the
  slow degrees of freedom, built on a corrotating frame.

\item[ii)] {\em Binaries and hellical  Killing vectors}.
  A most interesting construction for the dynamics of the
  slow degrees in  our present BBH setting is provided by
  Klein's hellical Killing model for quasi-circular binaries
  introduced in~\cite{Klein04,Klein:2006fdg}. In this case, a
  single hellical Killing vector (see e.g.~\cite{FriUryShi02}
  for a discussion of hellical Killing vectors) is imposed, instead
  of the two infinitesimal isometries leading to the Ernst equation discussed
  above. Applying again Ehlers \& Geroch's projection formalism,
  now only along one Killing vector, the four-dimensional Einstein equations reduce
  to three-dimensional gravity coupled to ``effective matter''
  described by a $SL(2, \mathbb{R}/SO(1,1)$ sigma model. The
  latter is determined by generalised Ernst-like equation (again for a
  complex potential, but living in the a three-dimensional quotient
  space).
  Even though the resulting system is not integrable as the standard (axisymmetric-stationary)
  Ernst equation is,  it provides a most interesting 
  starting point to explore the application of asymptotic methods
  leading to effective (quasi-)integrable dynamics for the background slow degrees
  of freedom.
  
\item[iii)] {\em Ward conjecture and BBH integrability}.
  A most remarkably connection between self-dual Yang Mills equations and
  integrable equations opens the possibility of an avenue to the use of twistorial techniques
  to address out BBH problem. More specifically, in a first step, a
  connection between the Ernst equation above in stationary axisymmetric spacetimes
  and certain self-dual Yang mills equations was identified by L. Witten in \cite{Witte79}.
  In a secone step, this was extended by Ward \cite{Ward83} to a more systematic and extensive treatment
  of the starionary axisymmetric case using twistor techniques (see also \cite{WooMas88}).
  This twistor insight into integrability in the key stationary axisymmetric situation in general relativity
  is already of major interest. However, this connection upgrades to a truly fundamental
  general connection in the perspective of Ward's conjecture~\cite{Ward85}, according to which
  {\em all integrable} (or solvable) differential equation systems can be obtained from reductions
  of self-dual Yang Mills equations.  Ref.~\cite{Ward85}
  raises a critical view on the use of the inverse scattering method to characterize integrability,
  since such method has limited applicability. On the contrary, the emphasis is placed
  in the satisfaction of the ``Painlev\'e property'' (solutions in terms of meromorphic functions)
  and the fact that integrable equations do appear as ``consitency conditions'' of an overdetermined
  system of {\em linear} PDEs (point shared with the inverse scattering method).
  Self-dual Yang Mills equations would then play the role of a sort of master
  equations, from which all integrable equations could be derived (see also
  \cite{Mason93,MasWoo96,AblChaHal03,Dunajski:2010zz}).
  A most remarkable fact in our gravity setting is then provided by the connection
  unveiled in \cite{MasNew89} between self-dual Einstein vacuum equations and the self-dual
  Yang-Mills equations. Under the light of Ward's conjecture, this suggests to 
  approach the identification
  of the integrable PDE describing the effective dynamics of the slow background degrees of motion
  in terms of self-dual Einstein vacuum equations and to incorporate the use of twistor concepts
  and tools to address the universality and simplicity of the BBH merger waveform.

\end{itemize}

\bigskip
 
\noindent{\em Acknowledgments.} We would like to acknowlegde Oleg Lisovyy and Peter Miller
for the key references that triggered our interest on Painlev\'e transcendents in the
BBH problem. We thank Abhay Ashtekar, Ivan Booth and Andrey Shoom for discussions
at the early stage of this project, at the ``Focus Session: Dynamical Horizons,
Binary Coalescences, Simulations and Waveform'' (State College, July 2018).
We thank  V\'\i ctor Aldaya, Carlos Barcel\'o,
Beatrice Bonga, A. Coutant, Christian Klein, Pavao Mardesic, O. Meneses-Rojas, Alex Nielsen,
Ariadna Ribes Metidieri, Dhruv Sharma and  Nikola Stoilov for discussions.
Special thanks to Carlos F. Sopuerta and Mikhail Semenov Tian-Shansky for enlightening
discussions on the overall project.

\bibliography{Biblio}

\end{document}